\documentclass{lmcs}


\usepackage{hyperref}

\usepackage[utf8]{inputenc}
\usepackage[vlined]{algorithm2e}
\usepackage{dafny}
\usepackage{tikz}
\usepackage{multirow}

\lstnewenvironment{datastructure}[1][]{%
  \lstset{language=dafny,#1}}
  {}

\newcommand{\Csh}{C\#}

\begin{document}

\title[A Computer-Checked Implementation of the DPLL Algorithm in
Dafny]{Who Verifies the Verifiers? A Computer-Checked Implementation
  of the DPLL Algorithm in Dafny\rsuper*}
  
\titlecomment{{\lsuper*}This is an extended version of a
  paper~\cite{EPTCS303.1} presented at the Working Formal Methods
  Symposium 2019.}

\author{Cezar-Constantin Andrici}
\address{Alexandru Ioan Cuza University, Iași}
\email{cezar.andrici@gmail.com}

\author{Ștefan Ciobâcă}
\address{Alexandru Ioan Cuza University, Iași}
\email{stefan.ciobaca@info.uaic.ro}

\begin{abstract}

  We build a SAT solver implementing the DPLL algorithm in the
  verification-enabled programming language Dafny. The resulting
  solver is fully verified (soundness, completeness and termination
  are computer checked). We benchmark our Dafny solver and we show
  that it is just as efficient as an equivalent DPLL solver
  implemented in C\# and roughly two times less efficient than an
  equivalent solver written in C++. We conclude that auto-active
  verification is a promising approach to increasing trust in SAT
  solvers, as it combines a good trade-off between execution speed and
  degree of trustworthiness of the final product.

\end{abstract}

\maketitle

\section{Introduction}

Modern high-performance SAT solvers quickly solve large satisfiability
instances that occur in practice. If the instance is satisfiable, then
the SAT solver can provide a witness in the form of a satisfying truth
assignment, which can be checked independently.

If the instance is unsatisfiable, the situation is less
clear. Brummayer and others~\cite{DBLP:conf/sat/BrummayerLB10} have
shown using fuzz testing that in 2010 many state-of-the-art SAT
solvers contained bugs, including soundness bugs. Since 2016, in order
to mitigate this issue, the annual SAT competition requires solvers
competing in the main track to output UNSAT
certificates~\cite{DBLP:conf/aaai/BalyoHJ17}; these certificates are
independently checked in order to ensure soundness.

These certificates could be exponentially large and the SAT solver
might not even be able to output them due to various resource
constraints. The implementation of the SAT solver should then be
trusted not to contain bugs. However, typical high-performance SAT
solvers contain data structures and algorithms complex enough to allow
for subtle programming errors.

To handle these potential issues, we propose a \emph{verified SAT
  solver} using the Dafny system~\cite{DBLP:conf/icse/Leino04}. Dafny
is a high-level imperative language with support for object oriented
features. It features methods with preconditions, postconditions and
invariants, which are checked at compilation time by relying on the Z3
SMT solver~\cite{DBLP:conf/tacas/MouraB08}. If a postcondition cannot
be established (either due to a timeout or due to the fact that it
does not hold), compilation fails. Therefore, we can place a high
degree of trust in a program verified using the Dafny system.

A modern high-performance SAT solver implements a backtracking search
for a satisfying assignment. The search space is pruned by using
several algorithmic tricks:
\begin{enumerate*}
\item unit propagation;
\item fast data structures (e.g., to identify unit clauses);
\item variable ordering heuristics;
\item back-jumping;
\item conflict analysis;
\item clause learning;
\item restart strategy.
\end{enumerate*}
\noindent In addition, careful engineering of the implementation is
required for high performance.

The first three items are usually referred to as the DPLL
algorithm~\cite{DBLP:journals/jacm/DavisP60,DBLP:journals/cacm/DavisLL62},
and all items together are the core of the state-of-the-art CDCL
algorithm~\cite{DBLP:journals/tc/Marques-SilvaS99,DBLP:conf/aaai/BayardoS97}. We
have implemented and verified in Dafny the first three items,
constituting the DPLL algorithm, and we leave the other items for
future work. We implement the MOMS variable ordering
heuristic~\cite{DBLP:journals/jar/HookerV95}. We note that our Dafny
solver is computer checked for soundness, completeness and
termination. We assume that the input is already in CNF form. The
parser, which reads a file in the well-known DIMACS format, is also
written in Dafny and hence verified against, e.g., out of bounds
errors. However, there is no specification for the parser.

Our work is part of the larger trend towards producing more
trustworthy software artifacts, ranging among certified
compilers~\cite{DBLP:journals/jar/Leroy09}, system
software~\cite{DBLP:journals/corr/abs-1004-3808,DBLP:conf/osdi/HawblitzelHLNPZZ14,DBLP:journals/ieeesp/BhargavanFK16,DBLP:conf/ccs/ZinzindohoueBPB17},
or logic~\cite{DBLP:journals/jar/BlanchetteFLW18}. The main conceptual
difference to previous work on verified or certified SAT solvers is
that we propose to check directly the imperative algorithm using
deductive verification, instead of, e.g., verifying functional code
and relying on a refinement mechanism to extract imperative code,
which could hurt performance.

\emph{Structure}. In Section~\ref{sec:dpll}, we briefly go over the
DPLL algorithm, as presented in the literature. In
Section~\ref{sec:verified}, we present our verified implementation in
Dafny of the algorithm. We start by presenting the main data
structures and their invariants (Section~\ref{subsec:ds}). We continue
with the operations supported by the data structures in
Section~\ref{subsec:op}. Finally, in Section~\ref{subsec:alg}, we
present the implementation of the core DPLL algorithm, together with
the verified guarantees that it provides.  In
Section~\ref{sec:benchmarks}, we benchmark the performance of our
solver. In Section~\ref{sec:related}, we discuss related work. We
conclude in Section~\ref{sec:conclusion}. We also discuss the main
challenge in verifying our implementation of DPLL, along with some
methodological tricks that we have used to make the verification
effort tractable.

\emph{Contributions}. We present the first (to our knowledge)
assertional proof of the DPLL algorithm. The implementation is
competitive in running time with an equivalent C++ solver.

\emph{Comparison with the workshop version}. This paper is a revised
extended version of our previous work~\cite{EPTCS303.1} published in
EPTCS. We feature an improved presentation, additional explanations
and a benchmark of the performance of our solver. In addition, the
solver improvements over the workshop version are:

\begin{enumerate}

\item The new implementation features machine integers, which improve
  performance approximately 10 times in our tests. Going to machine
  integers from unbounded integers requires proving upper bounds on
  indices throughout the code.

\item The new implementation features mutable data structures for
  identifying unit clauses. Our previous approach used Dafny sequences
  (\texttt{seq}), which are immutable and cause a performance drawback
  because they are updated frequently. The new mutable data structures
  make the solver significantly faster, but they are more difficult to
  reason about and verify.

\item We implement and verify the MOMS variable ordering heuristic.

\item We also improve the methodology of our verification approach and
  in particular we significantly reduce verification time. By
  carefully specifying invariants and separating concerns in the
  implementation, the verification time is now approximately 13
  minutes for the entire project.

  In contrast, in our previous implementation, one method
  (\texttt{setLiteral}) took approximately 10 minutes to verify on its
  own (the entire project used to take about 2 hours to verify in its
  entirety).

\item We benchmark our Dafny implementation against similar DPLL
  implementations written in C\# and C++ and we show it is competitive
  in terms of performance.
 
\end{enumerate}

\section{The Davis-Putnam-Logemann-Loveland Algorithm}
\label{sec:dpll}

The DPLL procedure is an optimization of backtracking search. The main
improvement is called unit propagation. A unit clause has the property
that its literals are all \textit{false} in the current assignment,
except one, which has no value yet. If this literal would be set to
\textit{false}, the clause would not be satisfied; therefore, the
literal must necessarily be \textit{true} for the entire formula to be
true. This process of identifying unit clauses and setting the unknown
literal to true is called unit propagation.

\begin{exa}\label{theexample} We consider a formula with 7 variables and 5 
clauses: 
\begin{center}
    $(x_{1} \lor x_{2} \lor x_{3}) \land$ 
    $(\lnot x_{1} \lor \lnot x_{2}) \land$
    $(x_{2} \lor \lnot x_{3}) \land $
    $(x_{2} \lor x_{4} \lor x_{5}) \land$
    $(x_{5} \lor x_{6} \lor x_{7})$
\end{center}
The formula is satisfiable, as witnessed by the truth assignment \textit{(true, false, false, true, true, false, true)}.
\end{exa}

Algorithm~\ref{alg:dpllrecursive} describes the DPLL
procedure~\cite{DBLP:reference/fai/GomesKSS08} that we implement and
verify, presented slightly differently in order to match our
implementation more closely:

\begin{algo}
  \label{alg:dpllrecursive}
\SetKwProg{main}{Function}{}{}
\SetKwFunction{funone}{DPLL-recursive}

\main{
    \funone{$F$, $\textit{tau}$}}{
    \SetKwInOut{Input}{input}\SetKwInOut{Output}{output}
    \Input{A CNF formula $F$ and an partial assignment $\textit{tau}$}
    \Output{SAT/UNSAT, depending on where there exists an assignment extending $\textit{tau}$ that satisfies $F$}
    \While{$\exists$ unit clause $\in F$}{
        $\ell \leftarrow$ the unset literal from the unit clause \\
        $\textit{tau} \leftarrow \textit{tau}[\ell := true]$
    }
    
    \lIf{F contains the empty clause}{\Return{UNSAT}}
    \If{F has no clauses left} {
        Output $\textit{tau}$ \\
        \Return{SAT}
    }
    
    $\ell \leftarrow$ some unset literal
    
    \lIf{$\textit{DPLL-recursive}(F, \textit{tau}[\ell := \textit{true}]) = \textit{SAT}$}{\Return{$\textit{SAT}$}}
    
    \Return{$\textit{DPLL-recursive}(F, \textit{tau}[\ell := \textit{false}])$}
}
\end{algo}

We describe how the algorithm works on this
example: first, the algorithm chooses the literal $x_{1}$ and sets it
to \textit{true} (arbitrarily; if \textit{true} would not work out,
then the algorithm would backtrack here and try \textit{false}). At
the next step, it finds that the second clause is unit and sets
$\lnot x_{2}$ to \textit{true}, which makes the third clause unit, so
$\lnot x_{3}$ is set to \textit{true}. After unit propagation, the
next clause not yet satisfied is the fourth one, and the first unset
literal is $x_{4}$. At the branching step, $x_{4}$ is assigned to
\textit{true}.  Furthermore, only one clause is not satisfied yet, and
the next decision is to choose $x_{5}$ and set it to \textit{true},
which makes the formula satisfied, even if $x_{6}$ and $x_{7}$ are not
set yet.

Next, we recall some well-known terminology in SAT solvers. Choosing
and assigning an unset literal to \textit{true} or \textit{false} is
called a branching step or a \emph{decision}. Every time the algorithm
makes a \emph{decision}, the decision level is incremented by one and
some more literals are assigned to \textit{true} or \textit{false} by
unit propagation.

The \textit{trace} of assignments is split into layers, one layer per
decision. Multiple literals can be set at the same decision level (the
decision literal, and the literals assigned by unit
propagation). Every time the algorithm backtracks it must revert an
entire layer of assignments. A possible assignments trace
corresponding to Example~\ref{theexample} is shown in
Figure~\ref{fig:stackpresentation}.

\begin{figure}[ht]
\begin{center}
\tikzstyle{clause} = [rectangle, text width=4cm, minimum height=0.45cm]
\tikzstyle{layer} = [rectangle, text width=6cm, minimum height=0.45cm, draw=black]
\begin{tikzpicture}[node distance=0.55cm]
\tikzstyle{every node}=[font=\small]
\node (clause1) [clause] {1) $\textcolor{green}{x_{1}} \lor \textcolor{red}{x_{2}} \lor \textcolor{red}{x_{3}}$};
\node (clause2) [clause, below of=clause1] {2) $\textcolor{red}{\lnot x_{1}} \lor \textcolor{green}{\lnot x_{2}}$};
\node (clause3) [clause, below of=clause2] {3) $\textcolor{red}{x_{2}} \lor \textcolor{green}{\lnot x_{3}}$};
\node (clause4) [clause, below of=clause3] {4) $\textcolor{red}{x_{2}} \lor \textcolor{green}{x_{4}} \lor x_{5}$};
\node (clause5) [clause, below of=clause4] {5) $\textcolor{green}{x_{5}} \lor x_{6} \lor x_{7}$};
\node (layer1) [layer, right of=clause1, xshift=5cm] {$(x_{1}, true)$, $(x_{2}, false)$, $(x_{3}, false)$};
\node (layer2) [layer, below of=layer1] {$(x_{4}, true)$};
\node (layer3) [layer, below of=layer2] {$(x_{5}, true)$};
\node (text1) [clause, above of=layer1, text width=6cm] {Example assignments trace:};
\node (text2) [clause, above of=clause1] {Example formula:};
\end{tikzpicture}\end{center}
\caption{\label{fig:stackpresentation}Assignments trace representation
  for Example~\ref{theexample} divided into layers. The first layer
  corresponds to setting the decision variable $x_{1}$ to
  $\textit{true}$, followed by unit propagation, which sets $x_{2}$
  and $x_{3}$. This trace occurs just before the algorithm stops with
  an answer of \texttt{SAT}, after setting the decision variables
  $x_4$ and $x_5$ to $\textit{true}$.}
\end{figure}

\section{A Verified Implementation of the DPLL Algorithm}

\label{sec:verified}

In this section, we present the main ingredients of our verified
solver. The full source code, along with instruction on how to compile
it and reproduce our benchmarks, can be found
at \begin{center}\url{https://github.com/andricicezar/sat-solver-dafny-v2}.\end{center}

\subsection{Data Structures}

\label{subsec:ds}

We first discuss the data structures for representing the formula, for
quickly identifying unit clauses and for recalling the current truth
assignment.

\subsubsection{Representing the CNF formula}

The main class in our Dafny development is \texttt{Formula}, which
extends \texttt{DataStructures} (Figure~\ref{fig:formulaDef}). This
class is instantiated with the number of propositional variables
(\texttt{variablesCount}) and with the \texttt{clauses} of the formula
to be checked for satisfiability. Propositional variables are
represented by values between $0$ and $\texttt{variablesCount}-1$,
positive literals are represented by values between $1$ and
\texttt{variablesCount}, and negative integers between $-1$ and
$-$\texttt{variablesCount} represent negative literals. Variables and
literals are represented by values of type \texttt{Int32.t}, which we
define to model machine integers and which is extracted to
\texttt{int}.

\begin{figure}[ht]
\begin{datastructure}
trait DataStructures {
    var variablesCount : Int32.t;
    var clauses : seq< seq<Int32.t> >;
    var decisionLevel : Int32.t;
...
    var traceVariable : array<Int32.t>;
    var traceValue : array<bool>;
    var traceDLStart : array<Int32.t>;
    var traceDLEnd : array<Int32.t>;  
    ghost var assignmentsTrace : set<(Int32.t, bool)>;
...
    var truthAssignment : array<Int32.t>;
    var trueLiteralsCount : array<Int32.t>;
    var falseLiteralsCount : array<Int32.t>;
    var positiveLiteralsToClauses : array< seq<Int32.t> >;
    var negativeLiteralsToClauses : array< seq<Int32.t> >;
}
\end{datastructure}
\caption{\label{fig:formulaDef}The most important fields in our data
  structures (file \texttt{solver/data\textunderscore structures.dfy}).}
\end{figure}

Clauses are sequences of literals and the entire formula is
represented by a sequence of clauses (\texttt{var clauses : seq<
  seq<Int32.t> >}). Using sequences for clause (sequences are
immutable in Dafny) has no significant performance impact, since they
are set at the beginning once and never changed.

\subsubsection{Representing the current assignment and the
  assignments trace}

The member variable \texttt{decisionLevel} recalls the current
decision level, which has an initial value of $-1$. The assignments
trace is represented at computation time by using the arrays
\texttt{traceVariable}, \texttt{traceValue}, \texttt{traceDLStart} and
\texttt{traceDLEnd} and at verification time also by the ghost
construct \texttt{assignmentsTrace} (see Figure~\ref{fig:formulaDef}).

The arrays \texttt{traceVariable} and \texttt{traceValue} have the
same actual length. They recall, in order, all variables that have
been set so far, together with their value. The arrays
\texttt{traceDLStart} and \texttt{traceDLEnd} recall at what index in
\texttt{traceVariable} and \texttt{traceValue} each decision layer
starts and ends, respectively.

The ghost construct \texttt{assignmentsTrace} recalls the same
information as a set of (variable, value) pairs. This set is used for
the convenience of specifying some of the methods and it only lives at
verification time; it is erased before running time and therefore it
entails no performance penalty.

Note that \texttt{traceVariable}, \texttt{traceValue},
\texttt{traceDLStart} and \texttt{traceDLEnd} are arrays, and they are
extracted to \Csh{} as such. Therefore, lookups and updates in these
arrays take constant time. The link between the ghost construct
\texttt{assignmentsTrace} and its imperative counterparts
(\texttt{traceVariable}, \texttt{traceValue}, \texttt{traceDLStart}
and \texttt{traceDLEnd}) is computer checked as the following class
invariant:

\begin{dafny}
(decisionLevel >= 0 ==> (
  (forall i :: 0 <= i < traceDLEnd[decisionLevel] ==> 
    (traceVariable[i], traceValue[i]) in assignmentsTrace)
  &&
  (forall x :: x in assignmentsTrace ==> (
    exists i :: 0 <= i < traceDLEnd[decisionLevel] &&
      (traceVariable[i], traceValue[i]) == x))))
\end{dafny}

The array \texttt{truthAssignment} is indexed from $0$ to
$\texttt{variablesCount}-1$ and it recalls the current truth
assignment. The value $\texttt{truthAssignment}[v]$ is $-1$ if the
propositional variable $v$ is unset, $0$ if $v$ is false, and $1$ if
$v$ is true. At the beginning, it is initialized to $-1$ at all
indices. The following class invariant describing the expected link
between the assignments trace and the current truth assignment is
computer checked:

\begin{dafny}
truthAssignment.Length == variablesCount &&
(forall i :: 0 <= i < variablesCount ==> -1 <= truthAssignment[i] <= 1) &&
(forall i :: 0 <= i < variablesCount && truthAssignment[i] != -1 ==>
  (i, truthAssignment[i]) in assignmentsTrace) &&
(forall i :: 0 <= i < variablesCount && truthAssignment[i] == -1 ==>
  (i, false) !in assignmentsTrace && (i, true) !in assignmentsTrace)
\end{dafny}

Note that the invariant makes use of the ghost construct
\texttt{assignmentsTrace} for brevity.

\subsubsection{Quickly identifying unit clauses}

The array \texttt{trueLiteralsCount} (\texttt{falseLiteralsCount}) is
used to recall how many literals in each clause are currently true
(resp. false). They are indexed from $0$ to
$|\texttt{clauses}|-1$. The value $\texttt{trueLiteralsCount}[i]$
denotes the number of literals set to true in $\texttt{clauses}[i]$
and $\texttt{falseLiteralsCount}[i]$ the number of false literals in
$\texttt{clauses}[i]$. These are used to quickly identify which
clauses are satisfied, which clauses are unit or which clauses are
false. For example, to check whether $\texttt{clauses}[i]$ is
satisfied, we simply evaluate $\texttt{trueLiteralsCount}[i] > 0$. The
following class invariant involving these arrays is computer checked:

\begin{dafny}
|trueLiteralsCount| == |clauses| &&
forall i :: 0 <= i < |clauses| ==>
  0 <= trueLiteralsCount[i] == countTrueLiterals(truthAssignment, clauses[i])
\end{dafny}
\noindent and analougously for \texttt{falseLiteralsCount}. Note that
\texttt{countTrueLiterals} is a function (not a method, hence it is
used for specification only) that actually computes the number of true
literals by walking through all literals in the respective clause.

In order to quickly update \texttt{trueLiteralsCount} and
\texttt{falseLiteralsCount} when a new literal is (un)set, we use
\texttt{positiveLiteralsToClauses} and
\texttt{negativeLiteralsToClauses}. These are arrays indexed from $0$ to
$\texttt{variablesCount}-1$. The first array contains the indices of
the clauses in which a given variable occurs. The second array
contains the indices of the clauses in which the negation of the given
variable occurs. They provably satisfy the following invariant:

\begin{dafny}
|positiveLiteralsToClauses| == variablesCount && (
  forall variable :: 0 <= variable < |positiveLiteralsToClauses| ==>
    ghost var s := positiveLiteralsToClauses[variable];
    ...
    (forall clauseIndex :: clauseIndex in s ==> variable+1 in clauses[clauseIndex]) &&
    (forall clauseIndex :: 0 <= clauseIndex < |clauses| && clauseIndex !in s ==>
      variable+1 !in clauses[clauseIndex]))
\end{dafny}
% "..." above
% replaces
% valuesBoundedBy(s, 0, |clauses|) && orderedAsc(s) &&
\noindent (analogously for $\texttt{negativeLiteralsToClauses}$).

To represent class invariants, Dafny encourages a methodology of
defining a class predicate \texttt{valid}. In our development,
\texttt{valid} consists of the conjunction of the above invariants,
plus several other lower-level predicates that we omit for
brevity. The predicate \texttt{valid} is used as a precondition and
postcondition for all class methods, and therefore plays the role of a
class invariant. This way, it is guaranteed that the data structures
are consistent.

\subsection{Verified Operations over the Data Structures}

\label{subsec:op}

From the initial (valid) state, we allow one of these four actions:
\begin{enumerate}
\item increase the decision level,
\item set a variable,
\item set a literal and perform unit propagation,
  and
\item revert the assignments done on the last decision
  level.
\end{enumerate}

Each of the actions is implemented as a method and we show that these
four methods preserve the data structure invariants above.

\subsubsection{The Method \texttt{increaseDecisionLevel}}

This method increments the decision level by one and creates a new
layer. The method guarantees that the new state is valid, and nothing
else changes. Its signature and its specification are:
\begin{dafny}
method increaseDecisionLevel()
  requires validVariablesCount();
  requires validAssignmentTrace();
  requires decisionLevel < variablesCount - 1;
  requires decisionLevel >= 0 ==>
    traceDLStart[decisionLevel] < traceDLEnd[decisionLevel];

  modifies `decisionLevel, traceDLStart, traceDLEnd;

  ensures decisionLevel == old(decisionLevel) + 1;
  ensures validAssignmentTrace();
  ensures traceDLStart[decisionLevel] == traceDLEnd[decisionLevel];
  ensures getDecisionLevel(decisionLevel) == {};
  ensures forall i :: 0 <= i < decisionLevel ==>
    old(getDecisionLevel(i)) == getDecisionLevel(i);
\end{dafny}

The predicates \texttt{validVariablesCount} and
\texttt{validAssignmentTrace} are used as conjuncts in the class
invariant. The function \texttt{getDecisionLevel} returns all
assignments at a given decision level as a set.

\subsubsection{The Method \texttt{setVariable}}

This method takes a variable that is not yet set and it updates is
value. Because the trace of assignments and \texttt{truthAssignment}
are changed, \texttt{trueLiteralsCount} and
\texttt{falseLiteralsCount} have to be updated. We use the arrays
\texttt{positiveLiteralsToClauses} and
\texttt{negativeLiteralsToClauses} to efficiently update them, and
prove that the clauses that are not mentioned in these arrays are not
impacted. The signature of \texttt{setVariable} and its specification
are:
\begin{dafny}
method setVariable(variable : Int32.t, value : bool)
  requires valid();
  requires validVariable(variable);
  requires truthAssignment[variable] == -1;
  requires 0 <= decisionLevel;

  modifies truthAssignment, traceVariable, traceValue,
           traceDLEnd, `assignmentsTrace, trueLiteralsCount,
           falseLiteralsCount;

  ensures valid();
  ensures traceDLStart[decisionLevel] < traceDLEnd[decisionLevel];
  ensures traceVariable[traceDLEnd[decisionLevel]-1] == variable;
  ensures traceValue[traceDLEnd[decisionLevel]-1] == value;

  // post conditions that ensure that only a position of the arrays
  // has been updated.
  ensures value == false ==> old(truthAssignment[..])[variable := 0]
    == truthAssignment[..];
  ensures value == true ==> old(truthAssignment[..])[variable := 1]
    == truthAssignment[..];
  ensures forall i :: 0 <= i < variablesCount && i != decisionLevel ==>
    traceDLEnd[i] == old(traceDLEnd[i]);
  ensures forall i :: 0 <= i < variablesCount && i != old(traceDLEnd[decisionLevel]) ==>
    traceVariable[i] == old(traceVariable[i]) && traceValue[i] == old(traceValue[i]);
  ensures forall x :: 0 <= x < old(traceDLEnd[decisionLevel]) ==>
    traceVariable[x] == old(traceVariable[x]);

  ensures forall i :: 0 <= i < decisionLevel ==>
    old(getDecisionLevel(i)) == getDecisionLevel(i);

  ensures assignmentsTrace == old(assignmentsTrace) + { (variable, value) };
 
  ensures countUnsetVariables(truthAssignment[..]) + 1 ==
    old(countUnsetVariables(truthAssignment[..]));
\end{dafny}      

\subsubsection{The Method \texttt{setLiteral}}

This method uses \texttt{setVariable} as a primitive, so the
preconditions and postconditions are similar. The main difference is
that after it makes the first update, it also performs unit
propagation, possibly recursively. This means that it calls
\texttt{setLiteral} again with new values. So, at the end of a call,
\texttt{truthAssignment} might change at several positions. To prove
termination, we use as a variant the number of unset variables, which
provably decreases at every recursive step. Its signature and its
specification are:
\begin{dafny}
method setLiteral(literal : Int32.t, value : bool)
  requires valid();
  requires validLiteral(literal);
  requires getLiteralValue(truthAssignment[..], literal) == -1;
  requires 0 <= decisionLevel;

  modifies truthAssignment, trueLiteralsCount,
           falseLiteralsCount, traceDLEnd, traceValue,
           traceVariable, `assignmentsTrace;

  ensures valid();
  ensures traceDLStart[decisionLevel] < traceDLEnd[decisionLevel];
  ensures forall x :: 0 <= x < old(traceDLEnd[decisionLevel]) ==>
    traceVariable[x] == old(traceVariable[x]);
  ensures assignmentsTrace == old(assignmentsTrace) +
    getDecisionLevel(decisionLevel);
  ensures forall i :: 0 <= i < decisionLevel ==>
    old(getDecisionLevel(i)) == getDecisionLevel(i);
  ensures countUnsetVariables(truthAssignment[..]) <
    old(countUnsetVariables(truthAssignment[..]));
  ensures (
    ghost var (variable, val) := convertLVtoVI(literal, value);
    isSatisfiableExtend(old(truthAssignment[..])[variable as int := val]) <==>
            isSatisfiableExtend(truthAssignment[..])
  );

  decreases countUnsetVariables(truthAssignment[..]), 0;
\end{dafny}

In the code above, the function
$\texttt{getLiteralValue}(\texttt{tau}, \ell)$ returns the value of
the literal $\ell$ in the truth assignment $\texttt{tau}$. Note that
the variable \texttt{truthAssignment} is an array, while
\texttt{truthAssignment[..]} converts the array to a sequence. The
sequence (immutable) is used to represent truth assignments at
specification level.

\subsubsection{The Method \texttt{revertLastDecisionLevel}}

This method reverts the assignments from in the last layer by changing
the value of the respective literals to $-1$. The proof of this method
requires several helper proofs that confirm that the data structures
are updated correctly. To quickly update \texttt{trueLiteralsCount}
and \texttt{falseLiteralsCount}, we again use the two arrays
\texttt{positiveLiteralsToClauses} and
\texttt{negativeLiteralsToClauses}. As part of postcondition, we prove
that the literals not on the last decision level remain unchanged:
\begin{dafny}
method revertLastDecisionLevel()
  requires valid();
  requires 0 <= decisionLevel;

  modifies `assignmentsTrace, `decisionLevel, truthAssignment, trueLiteralsCount,
           falseLiteralsCount, traceDLEnd;

  ensures decisionLevel == old(decisionLevel) - 1;
  ensures assignmentsTrace == old(assignmentsTrace) -
    old(getDecisionLevel(decisionLevel));
  ensures valid();
  ensures forall i :: 0 <= i <= decisionLevel ==>
    old(getDecisionLevel(i)) == getDecisionLevel(i);
  ensures decisionLevel > -1 ==>
    traceDLStart[decisionLevel] < traceDLEnd[decisionLevel];
\end{dafny}

\subsection{Proof of Functional Correctness for the Main Algorithm}

\label{subsec:alg}

The entry point called to solve the SAT instance is \texttt{solve}:

\begin{dafny}
method solve() returns (result : SAT_UNSAT)
  requires formula.valid();
  requires formula.decisionLevel > -1 ==>
    formula.traceDLStart[formula.decisionLevel] <
      formula.traceDLEnd[formula.decisionLevel];

  modifies formula.truthAssignment, formula.traceVariable, formula.traceValue,
           formula.traceDLStart, formula.traceDLEnd, formula`decisionLevel,
           formula`assignmentsTrace, formula.trueLiteralsCount,
           formula.falseLiteralsCount;

  ensures formula.valid();
  ensures old(formula.decisionLevel) == formula.decisionLevel;
  ensures old(formula.assignmentsTrace) == formula.assignmentsTrace;
  ensures forall i :: 0 <= i <= formula.decisionLevel ==>
    old(formula.getDecisionLevel(i)) == formula.getDecisionLevel(i);
  ensures formula.decisionLevel > -1 ==>
    formula.traceDLStart[formula.decisionLevel] <
      formula.traceDLEnd[formula.decisionLevel];

  ensures result.SAT? ==> formula.validValuesTruthAssignment(result.tau);  
  ensures formula.countUnsetVariables(formula.truthAssignment[..]) ==
    formula.countUnsetVariables(old(formula.truthAssignment[..]));
    
  ensures result.SAT? ==>  
    formula.isSatisfiableExtend(formula.truthAssignment[..]);
  ensures result.UNSAT? ==>
    !formula.isSatisfiableExtend(formula.truthAssignment[..]);

  decreases formula.countUnsetVariables(formula.truthAssignment[..]), 1;
\end{dafny}

It implements the $\textit{DPLL-procedure}$ given in
Algorithm~\ref{alg:dpllrecursive} using recursion. However, for
efficiency, the data structures are kept in the instance of a class
instead of being passed as arguments. The most important
postconditions stating the functional correctness are: if it returns
SAT then the current \texttt{truthAssignment} can be extended to
satisfy the formula, and if returns UNSAT it means that no truth
assignment extending the current \texttt{truthAssignment} satisfies
it. We use the predicate
$\texttt{isSatisfiableExtend}(\texttt{tau}, \texttt{clauses})$, which
tests whether there exists a complete assignment that extends the
partial truth assignment $\texttt{tau}$ and that satisfies the
formula.

We also show as a postcondition that \texttt{solve} ends in the same
state as where it starts. This means that we chose to undo the changes
even if we find a solution. Otherwise, the preconditions and postconditions
for solve would need to change accordingly and become more verbose and
less elegant. For simplicity, we chose to revert to the initial state
every time.

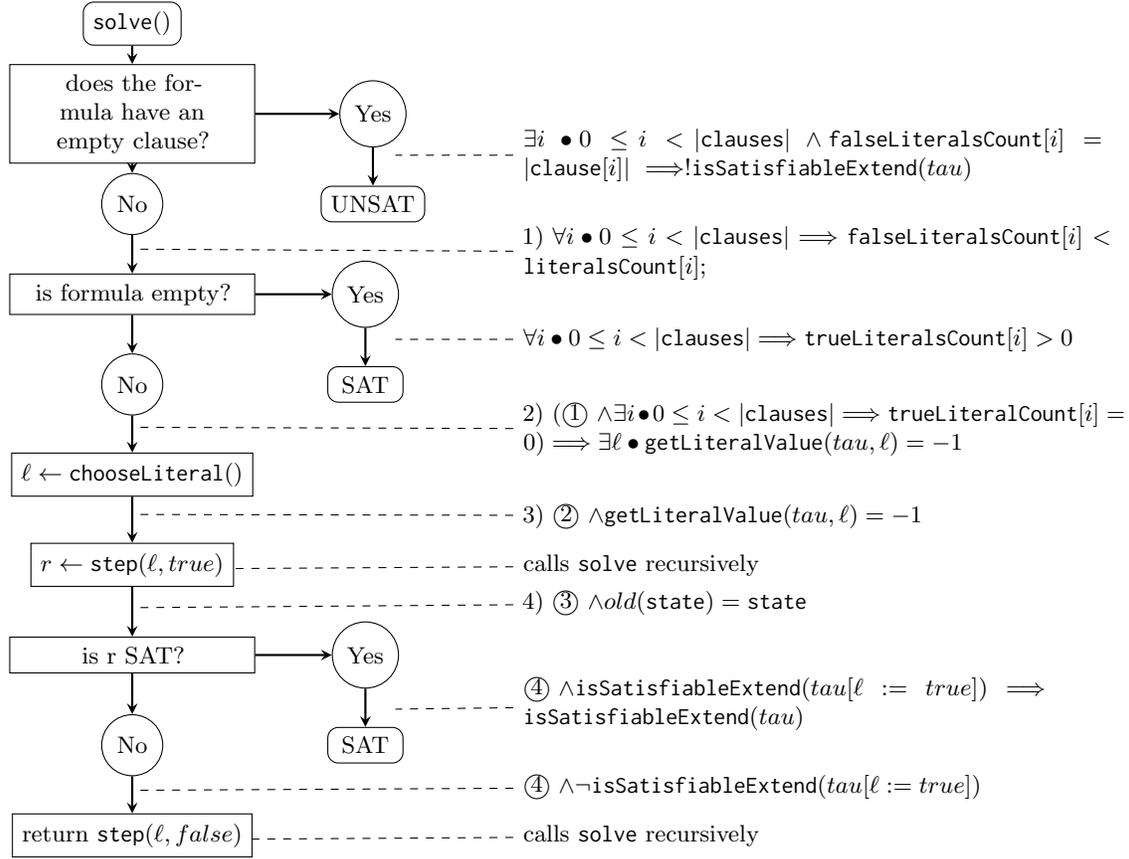
\begin{figure}[ht]
\tikzstyle{startstop} = [rectangle, rounded corners, minimum width=1cm, minimum height=0.1cm,text centered, draw=black]
\tikzstyle{command} = [rectangle, minimum height=0.1cm,text centered, draw=black ]
\tikzstyle{condition} = [rectangle, text width=3cm, text centered, draw=black]
\tikzstyle{conditionanswer} = [circle, text centered, draw=black]
\tikzstyle{proposition} = [rectangle, minimum height=0.1cm, text width=8cm]
\tikzstyle{arrow} = [thick,->,>=stealth]
\tikzstyle{arrow2} = [dashed,-,>=stealth]

\begin{tikzpicture}[node distance=1.2cm]
\tikzstyle{every node}=[font=\footnotesize]
\node (start) [startstop] {\texttt{solve}()};
\node (qhasEmptyClauses) [condition, below of=start] {does the formula have an empty clause?}; 
\draw [arrow] (start) -- (qhasEmptyClauses);
\node (qhasEmptyClausesNo) [conditionanswer, below of=qhasEmptyClauses] {No};
 \draw [arrow] (qhasEmptyClauses) -- (qhasEmptyClausesNo);
\node (qhasEmptyClausesYes) [conditionanswer, right of=qhasEmptyClauses, xshift=2cm] {Yes};
\draw [arrow] (qhasEmptyClauses) -- (qhasEmptyClausesYes);
\node (firstunsat) [startstop, below of=qhasEmptyClausesYes] {UNSAT};
\draw [arrow] (qhasEmptyClausesYes) -- (firstunsat);

\node (qisFormulaEmpty) [condition, below of=qhasEmptyClausesNo] {is formula empty?};
\draw [arrow] (qhasEmptyClausesNo) -- (qisFormulaEmpty);
\node (qisFormulaEmptyNo) [conditionanswer, below of=qisFormulaEmpty] {No};
\draw [arrow] (qisFormulaEmpty) -- (qisFormulaEmptyNo);
\node (qisFormulaEmptyYes) [conditionanswer, right of=qisFormulaEmpty, xshift=1.9cm] {Yes};
\draw [arrow] (qisFormulaEmpty) -- (qisFormulaEmptyYes);
\node (firstsat) [startstop, below of=qisFormulaEmptyYes] {SAT};
\draw [arrow] (qisFormulaEmptyYes) -- (firstsat);

\node (lchooseLiteral) [command, below of=qisFormulaEmptyNo] {$\ell \leftarrow \texttt{chooseLiteral}()$};
\draw [arrow] (qisFormulaEmptyNo) -- (lchooseLiteral);

\node (rstepltrue) [command, below of=lchooseLiteral] {$r \leftarrow \texttt{step}(\ell, true)$};
\draw [arrow] (lchooseLiteral) -- (rstepltrue);

\node (qisrsat) [condition, below of=rstepltrue] {is r SAT?};
\draw [arrow] (rstepltrue) -- (qisrsat);

\node (qisrsatNo) [conditionanswer, below of=qisrsat] {No};
\draw [arrow] (qisrsat) -- (qisrsatNo);
\node (qisrsatYes) [conditionanswer, right of=qisrsat, xshift=1.9cm] {Yes};
\draw [arrow] (qisrsat) -- (qisrsatYes);
\node (scndsat) [startstop, below of=qisrsatYes] {SAT};
\draw [arrow] (qisrsatYes) -- (scndsat);

\node (returnn) [command, below of=qisrsatNo] {return $\texttt{step}(\ell, false)$};
\draw [arrow] (qisrsatNo) -- (returnn);

\node (propunsat) [proposition, right of=firstunsat, yshift=0.65cm, xshift=4.8cm] {$\exists i\ \bullet 0\ \leq i\ < |\texttt{clauses}|\ \land \texttt{falseLiteralsCount}[i] \ =|\texttt{clause}[i] |\ \Longrightarrow !\texttt{isSatisfiableExtend}(tau)$};
\draw [arrow2] (propunsat) -- (3.5, -1.75);

\node (prop1) [proposition, right of=qhasEmptyClausesNo, yshift=-0.65cm, xshift=8cm] {1) $\forall i \bullet 0 \leq i < |\texttt{clauses}| \Longrightarrow \texttt{falseLiteralsCount}[i] < \texttt{literalsCount}[i];$};
\draw [arrow2] (prop1) -- (0, -3.00);

\node (propfstsat) [proposition, right of=firstsat, yshift=0.60cm, xshift=4.90cm] {$\forall i \bullet 0 \leq i < |\texttt{clauses}| \Longrightarrow \texttt{trueLiteralsCount}[i] > 0$};
\draw [arrow2] (propfstsat) -- (3.5, -4.20);

\node (prop2) [proposition, right of=qisFormulaEmptyNo, yshift=-0.60cm, xshift=8cm] {2) (\textcircled{1} $ \land \exists i \bullet 0 \leq i < |\texttt{clauses}| \Longrightarrow \texttt{trueLiteralCount}[i] = 0) \Longrightarrow \exists \ell \bullet \texttt{getLiteralValue}(tau, \ell) = -1$};
\draw [arrow2] (prop2) -- (0, -5.37);

\node (prop3) [proposition, right of=lchooseLiteral, yshift=-0.55cm, xshift=8cm] {3) \textcircled{2} $ \land  \texttt{getLiteralValue}(tau, \ell) = -1$};
\draw [arrow2] (prop3) -- (0, -6.53);

\node (prop4) [proposition, right of=rstepltrue, yshift=-0.50cm, xshift=8cm] {4) \textcircled{3} $ \land  old(\texttt{state}) = \texttt{state}$};
\draw [arrow2] (prop4) -- (0, -7.77);

\node (prop100) [proposition, right of=rstepltrue, yshift=0, xshift=8cm] {calls \texttt{solve} recursively};
\draw [arrow2] (prop100) -- (1.30, -7.25);

\node (propsndsat) [proposition, right of=scndsat, yshift=0.55cm, xshift=4.9cm] {\textcircled{4} $ \land \texttt{isSatisfiableExtend}(tau[\ell := true]) \Longrightarrow \texttt{isSatisfiableExtend}(tau)$};
\draw [arrow2] (propsndsat) -- (3.5, -9.12);

\node (propfinal) [proposition, right of=qisrsatNo, yshift=-0.55cm, xshift=8cm] {\textcircled{4} $ \land \lnot \texttt{isSatisfiableExtend}(tau[\ell := true])$};
\draw [arrow2] (propfinal) -- (0, -10.20);

\node (prop110) [proposition, right of=returnn, yshift=0, xshift=8cm] {calls \texttt{solve} recursively};
\draw [arrow2] (prop110) -- (1.5, -10.85);

\end{tikzpicture}
\caption{\label{fig:flowchartsolve}Flowchart of method
  \texttt{solve}. For simplicity, when the initial state is reached,
  we use the notation $\texttt{state} = \texttt{old(state)}$.}
\end{figure}

A flowchart that shows graphically the main flow of the \texttt{solve}
method, together with the most important statements that hold after
each line, is presented in Figure~\ref{fig:flowchartsolve}.

Once a literal is chosen, the updates to the data structures are
delegated to the \texttt{step} method. This removes some duplication
in the code, but it also makes the verification take less time. The
preconditions and postconditions of \texttt{step} are the same as in
\texttt{solve}, but taking into account that \texttt{step}
additionally takes as arguments a literal and a desired value for this
literal. The method \texttt{step} calls \texttt{setLiteral} to set the
literal and perform unit propagation, and then calls \texttt{solve}
recursively:

\begin{dafny}
method step(literal : Int32.t, value : bool) returns (result : SAT_UNSAT)
  requires formula.valid();
  requires formula.decisionLevel < formula.variablesCount - 1;
  requires formula.decisionLevel > -1 ==>
    formula.traceDLStart[formula.decisionLevel] <
      formula.traceDLEnd[formula.decisionLevel];
  requires !formula.hasEmptyClause();
  requires !formula.isEmpty();
  requires formula.validLiteral(literal);
  requires formula.getLiteralValue(formula.truthAssignment[..], literal) == -1;

  modifies formula.truthAssignment, formula.traceVariable, formula.traceValue, 
           formula.traceDLStart, formula.traceDLEnd, formula`decisionLevel,
           formula`assignmentsTrace, formula.trueLiteralsCount,
           formula.falseLiteralsCount;
             
  ensures formula.valid();
  ensures old(formula.decisionLevel) == formula.decisionLevel;
  ensures old(formula.assignmentsTrace) == formula.assignmentsTrace;
  ensures forall i :: 0 <= i <= formula.decisionLevel ==>
    old(formula.getDecisionLevel(i)) == formula.getDecisionLevel(i);
  ensures formula.decisionLevel > -1 ==>
    formula.traceDLStart[formula.decisionLevel] <
      formula.traceDLEnd[formula.decisionLevel];

  ensures result.SAT? ==> formula.validValuesTruthAssignment(result.tau);
  ensures result.SAT? ==> (
    var (variable, val) := formula.convertLVtoVI(literal, value);
    formula.isSatisfiableExtend(formula.truthAssignment[..][variable := val]));

  ensures result.UNSAT? ==> (
    var (variable, val) := formula.convertLVtoVI(literal, value);
    !formula.isSatisfiableExtend(formula.truthAssignment[..][variable := val]));

  ensures formula.countUnsetVariables(formula.truthAssignment[..]) ==
    formula.countUnsetVariables(old(formula.truthAssignment[..]));

  decreases formula.countUnsetVariables(formula.truthAssignment[..]), 0;
\end{dafny}

\section{Benchmarks}
\label{sec:benchmarks}

Dafny code can be extracted to \Csh{}, and then compiled and executed
as regular \Csh{} code. In this section, we present the results
obtained by benchmarking the \Csh{} code extracted from our verified
solver (we refer to this code as the Dafny solver) to see how it
performs against other solvers.

\emph{Benchmark used}. For benchmarking, we use some of the tests in
SATLIB - Benchmark
Problems\footnote{\url{https://www.cs.ubc.ca/~hoos/SATLIB/benchm.html}}. We
select the sets \texttt{uf100}, \texttt{uuf100} up to \texttt{uf200},
\texttt{uuf200}. These sets of SAT problems all contain instances in
3-CNF, with \texttt{uf} denoting satisfiable instances and
\texttt{uuf} denoting unsatisfiable instances. The numbers in the
names (e.g., \texttt{100}, \texttt{200}) denote the number of
propositional variables. The number of clauses in each set is chosen
such that the problems sit at the satisfiability
threshold~\cite{DBLP:journals/ai/CrawfordA96}. We choose these sets of
SAT instances because they are small enough for DPLL to solve in
reasonable time, but big enough so that the search dominates the
execution time (and not, e.g., reading the input).

\emph{Benchmarking methodology}. We run the tests using the
benchmarking framework BenchExec \cite{DBLP:journals/sttt/BeyerLW19},
a solution that reliably measures and limits resource usage of the
benchmarked
tool\footnote{\url{https://github.com/sosy-lab/benchexec}}. We used
BenchExec to limit resource usage to set the following for each run:
time limit to 5000s, memory limit to 1024 MB, CPU core limit to 1. We
used a Intel Core i7-9700K CPU @ 3.60GHz machine (cores: 4, threads:
8, frequency: 4900 MHz, Turbo Boost: enabled; RAM: 8290 MB, Operating
System Linux-5.3.0-40-generic-x86\_64-with-Ubuntu-18.04-bionic, Dafny
2.3.0.10506, Mono JIT compiler 6.8.0.105, G++ 7.5.0).

\emph{Benchmark 1}. We first check whether the extracted code has any
added overhead compared to a implementation written directly in
\Csh{}. For this purpose, we write in \Csh{} a solver implementing the
same algorithm and data structures as the Dafny solver.

We find that there is a negligible overhead coming from the method we
use to read files in Dafny, and not from the extraction process
itself. In our results, the reading and parsing of the input file in
Dafny takes at least twice as long as in \Csh{}. On small inputs, the
\Csh{} solver therefore outperforms the Dafny solver. On larger
inputs, the performance is the same.

\emph{Benchmark 2}. The language \Csh{} is not popular in SAT solving,
with C++ being the language of choice because of
performance. Therefore, we implement the same DPLL algorithm directly
in C++. We benchmark our verified Dafny solver against the C++
implementation. The results show that the (unverified) C++ solver is
approximately twice as fast on large tests as our verified Dafny
solver.

\emph{Benchmark 3}. To put the performance of our verified Dafny
solver into context, we also benchmark against the solver MiniSat
\footnote{\url{http://minisat.se/}} (with the default settings). As
MiniSat implements the full CDCL algorithm, which can be exponentially
faster than DPLL, it outperforms our solver significantly. However,
the correctness guarantee offered by our verified solver is higher
than the unverified C code of MiniSat.

In Table~\ref{tab:time}, we summarize the running times of all solvers
on the respective sets of tests. We report the average running time,
the standard deviation and the sum over all running times for SAT
instances in each particular set of tests.

\begin{table}[th]
\footnotesize
\setlength\tabcolsep{0.05cm}
    \centering
    \begin{tabular}{ | r | r | r | r | r | r | r | r | r | r | r | r | r | }
                                                                                                                                                                                
    \hline
    & \multicolumn{3}{c}{Dafny SAT Solver}\vline & \multicolumn{3}{c}{C\# DPLL Solver}\vline & \multicolumn{3}{c}{C++ DPLL Solver}\vline & \multicolumn{3}{c}{MiniSat v2.2.0}\vline \\
    \hline
     & avg & sd & sum & avg & sd & sum & avg & sd & sum & avg & sd & sum \\
    \hline 
    uf100 & 0.07 & 0.02 & 72.72 & 0.05 & 0.02 & 55.75 & 0.02 & 0.01 & 21.73 & 0.00 & 0.00 & 2.70 \\
    uuf100 & 0.13 & 0.03 & 130.58 & 0.11 & 0.03 & 111.65 & 0.04 & 0.01 & 49.94 & 0.00 & 0.00 & 4.03 \\
    uf150 & 1.29 & 1.30 & 129.24 & 1.22 & 1.26 & 122.22 & 0.63 & 0.66 & 63.54 & 0.00 & 0.00 & 0.84 \\
    uuf150 & 3.48 & 1.56 & 348.44 & 3.34 & 1.51 & 334.52 & 1.76 & 0.79 & 176.18 & 0.01 & 0.00 & 1.87 \\
    uf175 & 6.90 & 6.90 & 690.60 & 6.88 & 6.87 & 688.59 & 3.50 & 3.50 & 350.75 & 0.02 & 0.02 & 2.58 \\
    uuf175 & 21.66 & 10.60 & 2166.94 & 21.80 & 10.67 & 2180.41 & 10.77 & 5.25 & 1077.03 & 0.05 & 0.02 & 5.13 \\
    uf200 & 43.22 & 39.99 & 4322.50 & 47.07 & 43.69 & 4707.89 & 21.92 & 20.35 & 2192.21 & 0.06 & 0.05 & 6.48 \\
    uuf200 & 110.64 & 48.33 & 10953.40 & 120.07 & 52.64 & 11887.17 & 55.98 & 22.55 & 5542.17 & 0.18 & 0.08 & 17.82 \\
    \hline
    \end{tabular}
    \caption{\label{tab:time}The CPU time required to solve each set of instances by each SAT Solver in seconds. The first three solvers are implemented by us.}
\end{table}

\begin{figure}[h]
\centering
\includegraphics[width=10cm]{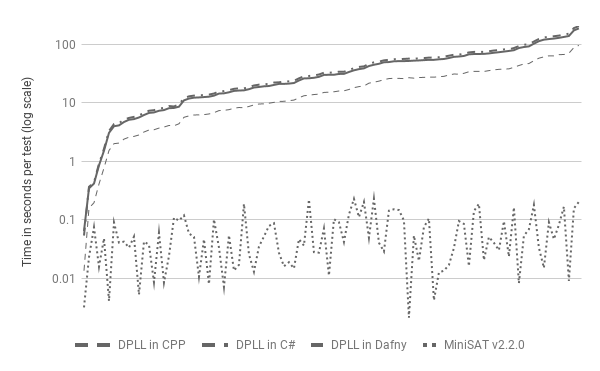}
\caption{\label{fig:uf200}CPU time on each instance in the set
  \texttt{uf200} (all instances are satisfiable in this set). The
  Dafny solver and the \Csh{} solver are essentially
  indistinguishable. The C++ solver is approximately twice as
  fast. MiniSAT is much faster, since it implements the full CDCL
  algorithm.}
\end{figure}

\begin{figure}[h]
\centering
\includegraphics[width=10cm]{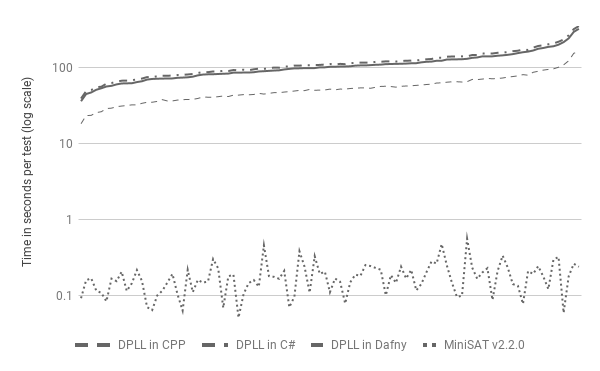}
\caption{\label{fig:uuf200}CPU time on each instance in the set
  \texttt{uuf200} (all instances are unsatisfiable in this set).  The
  Dafny solver and the \Csh{} solver are essentially
  indistinguishable. The C++ solver is approximately twice as
  fast. MiniSAT is much faster, since it implements the full CDCL
  algorithm.}
\end{figure}

Figures~\ref{fig:uf200} and~\ref{fig:uuf200} present the running times
(log scale) of all four solvers mentioned above on all SAT instances
in the \texttt{uf200} and \texttt{uuf200} sets, respectively. The
running times are sorted by the time it takes for the Dafny solver to
finish.

We conclude that our verified Dafny solver is competitive with an
equivalent implementation in C++ (it is only two times slower), but
the correctness guarantee offered by our verified solver makes it
significantly more trustworthy.

\section{Related Work}
\label{sec:related}

The SAT solver {\tt versat}~\cite{DBLP:conf/vmcai/OeSOC12} was
implemented and verified in the Guru programming language using
dependent types. As our solver, it also implements efficient data
structures. However, it relies on a translation to C where data
structures are implemented imperatively by using reference counting
and a statically enforced read/write discipline. Unlike our approach,
the solver is only verified to be sound: if it produces an {\tt UNSAT}
answer, then the input formula truly is unsatisfiable. However,
termination and completeness (if the solver produces SAT, then the
formula truly is satisfiable) are not verified. Another small
difference is the verification guarantee: {\tt versat} is verified to
output {\tt UNSAT} only if a resolution proof of the empty clause
exists, while in our approach we use a semantic criterion: our solver
always terminates and produces UNSAT only if there is no satisfying
model of the input formula. Of course, in the case of propositional
logic these criteria are equivalent and therefore this difference is
mostly a matter of implementation. Unlike our solver, some checks are
not proved statically and must be checked dynamically, so they could
be a source of incompleteness. An advantage of {\tt versat} over our
approach is that is implements more optimizations, like conflict
analysis and clause learning, which enable it to be more competitive
in terms of running time. Blanchette and
others~\cite{DBLP:journals/jar/BlanchetteFLW18} present a certified
SAT solving framework verified in the Isabelle/HOL proof
assistant. The proof effort is part of the \emph{Isabelle
  Formalization of Logic} project. The framework is based on
refinement: at the highest level sit several calculi like CDCL and
DPLL, which are formally proved. Depending on the strategy, the
calculi are also shown to be terminating. The calculi are shown to be
refined by a functional program. Finally, at the lowest level is an
imperative implementation in Standard ML, which is shown to be a
refinement of the functional implementation. Emphasis is also placed
on meta-theoretical consideration. The final solver can still two
orders of magnitude slower than a state-of-the-art C solver and
therefore additional optimizations~\cite{DBLP:conf/nfm/Fleury19} are
desirable. In contrast, in our own work we do not investigate any
meta-theoretical properties of the DPLL/CDCL frameworks; we simply
concentrate on obtaining a verified SAT solver. We investigate to what
extent directly proving the imperative algorithm is possible in an
auto-active manner. A key challenge is that the verification of Dafny
code may take a lot of time in certain cases and we have to optimize
our code for verification time as well. Another SAT solver verified in
Isabelle/HOL, is by Mari\'{c}~\cite{DBLP:journals/jar/Maric09}. In
contrast to previous formalization, the verification methodology is
not based on refinement. Instead, the Hoare triples associated to the
solver pseudo-code are verified in Isabelle/HOL. In subsequent
work~\cite{DBLP:journals/corr/abs-1108-4368}, Mari{\'c} and Janičić
prove in Isabelle the functional correctness of a SAT solver
represented as an abstract transition system. Another formalization of
a SAT solver (extended with linear arithmetic) is by
Lescuyer~\cite{lescuyer:tel-00713668}, who verifies a DPLL-based
decision procedure for propositional logic in Coq and exposes it as a
reflexive tactic. Finally, a decision procedure based on DPLL is also
verified by Shankar and Vaucher~\cite{DBLP:journals/entcs/ShankarV11}
in the PVS system. For the proof, they rely on subtyping and dependent
types. Berger et al. have used the Minlog proof assistant to extract a
certified SAT solver~\cite{DBLP:journals/corr/BergerLFS15}. For these
last approaches, performance considerations seem to be secondary.

\section{Conclusion and Further Work}
\label{sec:conclusion}

We have developed a formally verified implementation of the DPLL
algorithm in the Dafny programming language. Our implementation is
competitive in terms of execution time, but it is also trustworthy:
all specifications are computer checked by the Dafny system. Other
approaches to SAT solvers that rely on type
checkers~\cite{DBLP:journals/jar/BlanchetteFLW18} are arguably even
more trustworthy, since they are verified by a software system
satisfying the de Bruijn criterion. However, we believe that our
approach can strike a good balance between efficiency and
trustworthiness of the final product.

Our implementation incorporates data structures to quickly identify
unit clauses and perform unit propagation. The formalization consists
of around 3088 lines of Dafny code, including the parser. The code was
written by the first author in approximately one year and a half of
part time work. The author also learned Dafny during that time. The
ratio between lines of proof and lines of code is approximately
4/1. Table~\ref{tab:stats} contains a summary of our verified solver
in numbers.

\begin{table}[th!]
    \centering
    \begin{tabular}{ | p{3cm} | p{6.2cm} | p{3.9cm} | p{0.7cm} | }
      
      \hline
      Lines of code & 3088 (without whitespace) & Preconditions & 420 \\
      \hline
      Classes & 4 (and 1 trait) & Postconditions & 181 \\
      \hline
      Methods & 33 & Invariants & 173 \\
      \hline
      \multirow{3}{2cm}{Verification time} & approx. 13 minutes (entire project) & Variants & 44 \\
       & most expensive: \texttt{SATSolver.solve}

\qquad (approx. 175s)
& Predicates

Functions & 37 20 \\
       & more than 60s:

\qquad 6 methods/lemmas

between 10s and 60s:

\qquad 2 methods/lemmas & 
                           Ghost variables

Lemmas
    
                           Assertions & 24

42
    
    169 \\
    \hline
      Ratio specification/code & 2488 lines of specification/proofs to

600 lines of code & \texttt{reads} annotations

\texttt{modifies} annotations & 41 26 \\
    \hline
  \end{tabular}
  \caption{\label{tab:stats}Various statistics for our verified DPLL solver.}
\end{table}

In addition to coming up with the right invariants, the main challenge
in the development of the verified solver is the large amount of time
required by the Dafny system to discharge the verification
conditions. In order to minimize this verification time, we develop
and use the following development/verification methodology:

\begin{enumerate}

\item Avoid nested loops in methods. Nested loops usually require
  duplicating invariants, which decreases elegance and increases
  verification time.

  An example of applying this tip is the
  \texttt{revertLastDecisionLevel} method (in the file
  \texttt{solver/formula.dfy}), whose purpose is to backtrack to the
  previous decision level. The code of the method is currently very
  simple: it calls \texttt{removeLastVariable} repeatedly in a while
  loop. However, because it is so simple, it is tempting to inline
  \texttt{removeLastVariable} -- this would lead to a significant
  increase in verification time.

\item In the same spirit, avoid multiple quantifications in
  specifications.

  We have found it useful, whenever having a specification of the form
  \texttt{forall x :: forall y :: P(x, y)}, to try to extract the
  subformula \texttt{forall y :: P(x, y)} as a separate predicate of
  \texttt{x}. This helps in two distinct ways: it forces the
  programmer to name the subformula, thereby clearing their thought
  process and making their intention more clear, and it enables the Z3
  pattern-based quantifier instantiation to perform
  better~\cite{10.1145/1670412.1670416}.

\item Use very small methods. We find that it is better to extract as
  a method even code that is only a few lines of code long. In a usual
  programming language, such methods would be inlined (by the
  programmer). In our development, it is not unusual for such methods
  (with very few lines of code) to require many more helper
  annotations (invariants, helper assertions, etc.) and take
  significant time to verify.

\item Use minimal \texttt{modifies} clauses in methods and
  \texttt{reads} clauses in functions. In particular, we make
  extensive use of the less well-known backtick operator in Dafny.

\item During development, use Dafny to verify only the lemma/method
  currently being worked on. Run Dafny on the entire project at the
  end. To force Dafny to check only one method, we use the
  \texttt{-proc} command line switch.

\item Finally, we have found that using the rather nice Z3 axiom
  profiler~\cite{DBLP:conf/tacas/BeckerMS19} to optimize verification
  time does not scale well to projects the size of our solver.

\end{enumerate}

Our project shows that it is possible to obtain a fully verified SAT
solver written in assertional style, solver that is competitive in
terms of running time with similar solvers written in non-verifiable
languages. However, our experience with the verified implementation of
the solver is that it currently takes significant effort and expertise
to achieve this. We consider that three directions of action for the
development of Dafny (and other similar auto-active verification
tools) would be beneficial in order to improve this situation:

\begin{enumerate}
  
\item Improve verification time of individual methods/lemmas,

\item Make failures of verification obligations to check more
  explainable, and

\item Devise a method better than \texttt{assert}s to guide the
  verifier manually.

\end{enumerate}

As future work, we would like to verify an implementation of the full
CDCL algorithm, thereby obtaining a verified solver that is
competitive against state-of-the-art SAT solvers. In order to upgrade
to a competitive CDCL solver, we need to modify the algorithm to
implement a back-jumping and clause learning strategy, but also
implement the two watched literals data
structure~\cite{DBLP:reference/fai/GomesKSS08}, which becomes more
important for performance when the number of clauses grows.

\section*{Acknowledgments}

This work was supported by a grant of the Alexandru Ioan Cuza
University of Ia{\c s}i, within the Research Grants program UAIC
Grant, code GI-UAIC-2018-07.

\bibliographystyle{alpha}

\bibliography{submission}

\end{document}